\pdfoutput=1
\documentclass[12pt,a4paper]{article}
\usepackage{fullpage}
\usepackage[OT4]{fontenc}
\usepackage{amsopn}
\usepackage{amsmath}
\usepackage{amssymb}
\usepackage{amsthm}
\usepackage{hyperref}
\usepackage{array}
\usepackage{graphicx}
\usepackage{subfig}
\usepackage{color}
\usepackage{cite}
\newcommand{\ket}[1]{\ensuremath{|#1\rangle}}
\newcommand{\bra}[1]{\ensuremath{\langle#1|}}
\newcommand{\ketbra}[2]{\ensuremath{\ket{#1}\bra{#2}}}

\newcommand{\Tr}{\mathrm{Tr}}
\newcommand{\tr}{\Tr}
\newcommand{\1}{{\rm 1\hspace{-0.9mm}l}}

\title{Relativistic quantum pseudo-telepathy}

\author{P. Gawron  \quad {\L}. Pawela\\
Institute of Theoretical and Applied Informatics,\\
Polish Academy of Sciences,\\
Ba{\l}tycka 5, 44-100 Gliwice, Poland}

\date{09.IX.2014}

\begin{document}
\maketitle

\begin{abstract}
We analyze the impact of the Unruh effect on quantum Magic Square game. We find
the values of acceleration parameter for which quantum strategy yields higher
players' winning probability than classical strategy.

\end{abstract}
%%%%%%%%%%%%%%%%%%%%%%%%%%%%%%%%%%%%%%%%%%%%%%%%%%%%%%%%%%%%%%%%%%%%%%%%%%%%%%%%
\section{Introduction}\label{sec:introduction}
%%%%%%%%%%%%%%%%%%%%%%%%%%%%%%%%%%%%%%%%%%%%%%%%%%%%%%%%%%%%%%%%%%%%%%%%%%%%%%%%
Quantum game theory is an interdisciplinary field that combines game theory
and quantum information. It lies at the crossroads of physics, quantum
information processing, computer and natural sciences. Various quantizations of
games were presented by different authors 
\cite{eisert1999quantum,benjamin2000comment,piotrowski2003invitation}.

Quantum pseudo-telepathy games~\cite{brassard2005quantum} form a subclass of
quantum games. A game belongs to the pseudo-telepathy class providing that 
there are no winning strategies for classical players, but a winning strategy 
can be found if the players share a sufficient amount of entanglement. In these 
games quantum players can accomplish tasks that are unfeasible for their 
classical counterparts. 

Given a pseudo-telepathy game, one can implement a quantum winning strategy for
this game \cite{brassard2005quantum}. In an ideal case, the experiment should
involve a significant number of rounds of the game. The experiment should be
continued until either the players lose a single round or the players win such a
great number of rounds, that it would be nearly impossible if they were using a
classical strategy. Unfortunately ideal noiseless setup of a quantum game is
impossible to be achieved.

The motivation to study the Magic Square game and pseudo-telepathy games in
general is that their physical implementation could provide convincing, even to
a layperson, demonstration that the physical world is not local realistic. By
\emph{local} we mean that no action performed at some location X can have an
effect on some remote location Y in a time shorter then that required by light
to travel from X to Y. \emph{Realistic} means that a measurement can only
reveal elements of reality that are already present in the
system~\cite{brassard2005quantum}.

It has been shown~\cite{noisy_square} that noise in a 
quantum channel can decrease the probability of winning the Magic Square game 
even below the classical threshold, although the players can counteract this 
effect for lower noise~\cite{pawela2013enhancing}.
%%%%%%%%%%%%%%%%%%%%%%%%%%%%%%%%%%%%%%%%%%%%%%%%%%%%%%%%%%%%%%%%%%%%%%%%%%%%%%%%
\section{Magic Square game}\label{sec:square}
%%%%%%%%%%%%%%%%%%%%%%%%%%%%%%%%%%%%%%%%%%%%%%%%%%%%%%%%%%%%%%%%%%%%%%%%%%%%%%%%
The magic square is a $3 \times 3$ matrix filled with numbers 0 or 1 so that the
sum of entries in each row is even and the sum of entries in each column is odd.
Although such a matrix cannot exist (see Fig.~\ref{fig:impossible}) one can
consider the following game.

\begin{figure}[!htp]
	\centering
	\begin{tabular}{|c|c|c|}
	\hline
	1 & 1 & 0 \\ \hline
	1 & 0 & 1 \\ \hline
	1 & 0 & ? \\ \hline
	\end{tabular}
	\caption{An illustrative filling of the magic square with numbers 0 and 1. 
	The 
	question mark shows that it is not possible to put a number in the last 
	field and satisfy both conditions of the game.}\label{fig:impossible}
\end{figure}

The game setup is as follows. There are two players: Alice and Bob. Alice is
given a row, Bob is given a column. Alice has to give the entries for a row and
Bob has to give the entries for a column so that the parity conditions are met.
Winning condition is that the players' entries at the intersection must agree.
Alice and Bob can prepare a strategy but they are not allowed to communicate
during the game.

There exists a (classical) strategy that guarantees the winning probability of
$\frac{8}{9}$. If the parties are allowed to share a quantum state they can
achieve probability of success equal to one \cite{brassard2005quantum}.

In the quantum version of this game~\cite{mermin, aravind2004quantum} Alice and
Bob are allowed to share an entangled quantum state. The winning strategy is
following. Alice and Bob share entangled state
\begin{equation}
\ket{\psi} = \frac{1}{2}\left( \ket{0011} + \ket{1100} - \ket{0110} - 
\ket{1001} \right)\label{eq:init-state}
\end{equation}
and apply local unitary operators $A_i\otimes B_j$, where
\begin{center}
\begin{tabular}{l@{=}ll@{=}l}
$A_1$ &
$
\frac{1}{\sqrt{2}}
\left(
\begin{smallmatrix}
i & 0 & 0 & 1\\
0 &-i & 1 & 0\\
0 & i & 1 & 0\\
1 & 0 & 0 & i
\end{smallmatrix}
\right)
,
$ &
$A_2$ &
$
\frac{1}{2}
\left(
\begin{smallmatrix}
i & 1 & 1 & i\\
-i & 1 & -1 & i\\
i & 1 & -1 & -i\\
-i & 1 & 1 & -i
\end{smallmatrix}
\right)
,
$ \\
$A_3$ &
$
\frac{1}{2}
\left(
\begin{smallmatrix}
-1 & -1 & -1 & 1\\
1 & 1 & -1 & 1\\
1 & -1 & 1 & 1\\
1 & -1 & -1 & -1
\end{smallmatrix}
\right)
,
$ &
$B_1$ &
$
\frac{1}{2}
\left(
\begin{smallmatrix}
i & -i & 1 & 1\\
-i & -i & 1 & -1\\
1 & 1 & -i & i\\
-i & i & 1 & 1
\end{smallmatrix}
\right)
,
$ \\
$B_2$ &
$
\frac{1}{2}
\left(
\begin{smallmatrix}
-1 & i & 1 & i\\
1 & i & 1 & -i\\
1 & -i & 1 & i\\
-1 & -i & 1 & -i
\end{smallmatrix}
\right)
,
$ &
$B_3$&
$
\frac{1}{\sqrt{2}}
\left(
\begin{smallmatrix}
1 & 0 & 0 & 1\\
-1 & 0 & 0 & 1\\
0 & 1 & 1 & 0\\
0 & 1 & -1 & 0
\end{smallmatrix}
\right).
$
\end{tabular}
\end{center}
Indices $i$ and $j$ label rows and columns of the magic square.
Therefore the state of this scheme before measurement is
\begin{equation}\label{eq:game-final-state}
\rho_f=(A_i\otimes B_j)\,
\ket{\psi}\bra{\psi}\,(A_i^\dagger\otimes B_j^\dagger).
\end{equation}
The final
step of the game consists of the measurement in the computational basis.

We are interested in the mean probability of Alice and Bob winning the game. This 
probability is given by
\begin{equation}
p=\frac{1}{9}
\sum\limits_{i,j=1}^{3}
\sum_{\xi\in\mathcal{S}_{ij}}
\Tr{\rho_f}
{\ketbra{\xi}{\xi}},\label{eq:prob}
\end{equation}
where $\mathcal{S}_{ij}$ is the set of right answers for the column and row 
$ij$ (Tab.~\ref{tbl:msg:states}). In the case of no acceleration  we get $p=1$.

{
\renewcommand{\baselinestretch}{1}
\renewcommand{\arraystretch}{1}
\renewcommand{\tabcolsep}{0.4mm}
\begin{table}[h]
\begin{center}
\begin{footnotesize}
\begin{tabular}{c|cccccccccccccccc}
 & 
 ${0}$&${1}$&${2}$&${3}$&${4}$&${5}$&${6}$&${7}$
 &${8}$&${9}$&${10}$&${11}$&${12}$&${13}$&${14}$&
 ${15}$\\
\hline
$\mathcal{S}_{11}$&+&+&-&-&+&+&-&-&-&-&+&+&-&-&+&+\\
$\mathcal{S}_{12}$&+&+&-&-&-&-&+&+&+&+&-&-&-&-&+&+\\
$\mathcal{S}_{13}$&+&+&-&-&-&-&+&+&-&-&+&+&+&+&-&-\\
$\mathcal{S}_{21}$&+&-&+&-&+&-&+&-&-&+&-&+&-&+&-&+\\
$\mathcal{S}_{22}$&+&-&+&-&-&+&-&+&+&-&+&-&-&+&-&+\\
$\mathcal{S}_{23}$&+&-&+&-&-&+&-&+&-&+&-&+&+&-&+&-\\
$\mathcal{S}_{31}$&-&+&+&-&-&+&+&-&+&-&-&+&+&-&-&+\\
$\mathcal{S}_{32}$&-&+&+&-&+&-&-&+&-&+&+&-&+&-&-&+\\
$\mathcal{S}_{33}$&-&+&+&-&+&-&-&+&+&-&-&+&-&+&+&-
\end{tabular}
\end{footnotesize}
\end{center}
\caption{Sets $\mathcal{S}_{ij}$ --- plus sign (+) indicates that the given 
element belongs to the set, minus (-) sign indicates that the element does not 
belong to the set.}
\label{tbl:msg:states}
\end{table}
}
%%%%%%%%%%%%%%%%%%%%%%%%%%%%%%%%%%%%%%%%%%%%%%%%%%%%%%%%%%%%%%%%%%%%%%%%%%%%%%%%
\section{Magic square game in non-inertial reference frames}\label{sec:non-interial}
%%%%%%%%%%%%%%%%%%%%%%%%%%%%%%%%%%%%%%%%%%%%%%%%%%%%%%%%%%%%%%%%%%%%%%%%%%%%%%%%
In order to derive expressions for the Magic Square game in a non-inertial
reference frame, let us consider the initial state of the game as an entangled
state of four fermionic qubits of mode frequencies $\omega_\mathrm{A}$ and
$\omega_\mathrm{B}$ in a flat Minkowski space-time:
\begin{equation}
\ket{\psi} = \frac12 \left( \ket{0_{\omega_\mathrm{A}}} 
\ket{1_{\omega_\mathrm{B}}} - \ket{1_{\omega_\mathrm{A}}} 
\ket{0_{\omega_\mathrm{B}}}\right)
\otimes
\left( \ket{0_{\omega_\mathrm{A}}} \ket{1_{\omega_\mathrm{B}}} - 
\ket{1_{\omega_\mathrm{A}}} 
\ket{0_{\omega_\mathrm{B}}}\right),\label{eq:init-state-accel}
\end{equation}
where kets $\ket{0_{\omega_\mathrm{x}}}$ and $\ket{1_{\omega_\mathrm{x}}}$ 
($\mathrm{x} \in \{\mathrm{A}, \mathrm{B}\}$) represent the vacuum and excited 
states from the perspective of an inertial observer. The state in 
Eq.~\eqref{eq:init-state-accel} is equivalent to the one defined in 
Eq.~\eqref{eq:init-state} up to a swap operation.

In order to describe accelerated observers, we use Rindler coordinates, which 
define two casually disconnected regions ($I, II$), \textit{i.e.} a uniformly 
accelerated observer in region $I$ has no access to information in region $II$ 
and vice versa~\cite{alsing2006entanglement}. In order to simplify our scheme, 
we consider a single mode in the Rindler region $I$. This approach is justified 
if the observers' detectors are highly monochromatic and detect the frequency 
$\omega_\mathrm{A} \approx \omega_\mathrm{B} = \omega$. Hereafter, due to this 
approximation we will omit the subscript of $\omega$.

In an accelerated reference frame, the vacuum state becomes a two mode squeezed 
state given 
by~\cite{alsing2006entanglement,aspachs2010optimal,martin2010quantum,bruschi2010unruh}
\begin{equation}
\ket{0}=\cos(r)  \ket{0}_I\ket{0}_{II} + \sin(r) 
\ket{1}_I\ket{1}_{II},\label{eq:black-body}
\end{equation}
and the excited state becomes
\begin{equation}
\ket{1}=\ket{1}_I\ket{0}_{II},
\end{equation}
where $I$ and $II$ represents mode in the two Rindler regions. The parameter 
$r$ is a dimensionless acceleration given by:
\begin{equation}
r = \sqrt{\exp\left( \frac{-2\pi \omega c}{a} \right) + 1},
\end{equation}
where $a$ is Bob's acceleration. The  value $r=0$ corresponds to no
acceleration and the value $r=\frac{\pi}{4}$ corresponds to infinite
acceleration.

Eq.~\eqref{eq:black-body} shows that an observer in an non-inertial reference 
frame, moving with a constant acceleration, sees a thermal state instead of a 
vacuum state, \textit{i.e.} observers a black body radiation. This phenomenon is called 
the Unruh effect~\cite{unruh1976notes,davies1975scalar}

In the case of stationary Alice and accelerating Bob, using the Rindler 
coordinates we obtain obtain the following initial state
\begin{equation}
\ket{\psi} = \ket{\psi_{(1)}} \otimes \ket{\psi_{(2)}},
\end{equation}
where
\begin{equation}
\begin{split}
\ket{\psi_{(i)}}& = 
\frac12 ( 
\ket{0_{I\mathrm{A}_{(i)}}}\ket{0_{II\mathrm{A}_{(i)}}}\ket{1_{I\mathrm{B}_{(i)}}}\ket{0_{II\mathrm{B}_{(i)}}} -
\\& - 
\cos(r)\ket{1_{I\mathrm{A}_{(i)}}}\ket{0_{II\mathrm{A}_{(i)}}}\ket{0_{I\mathrm{B}_{(i)}}}\ket{0_{II\mathrm{B}_{(i)}}}
- 
\sin(r)\ket{1_{I\mathrm{A}_{(i)}}}\ket{0_{II\mathrm{A}_{(i)}}}\ket{1_{I\mathrm{B}_{(i)}}}\ket{1_{II\mathrm{B}_{(i)}}}
),
\end{split}
\end{equation}
which follows from using $r=0$ for Alice's qubits.

As there is no access to information in region $II$, we trace over modes in 
this region. Then we apply partial swap operation $U_\mathrm{SWAP}$. We get the 
following density matrix:
\begin{equation}
\rho_r = U_\mathrm{SWAP}
\tr_{II\mathrm{A}_{(1)} II\mathrm{B}_{(1)} II\mathrm{A}_{(2)} 
II\mathrm{A}_{(2)}}(\ketbra{\psi}{\psi})U_\mathrm{SWAP}^\dagger,\label{eq:rho-r}
\end{equation}
where $U_\mathrm{SWAP}$ is given by
\begin{equation}
U_\mathrm{SWAP} = \1_2 \otimes \left( \sum_{(i, j) \in \{0, 1\}^{\times 2}} 
\ketbra{ji}{ij} \right)\otimes \1_2.
\end{equation}
Note that for $r=0$ we recover the state from Eq.~\eqref{eq:init-state}.

The final state of the game depending on question pair $(i, j)$ is:
\begin{equation}\label{eq:game-final-noisy}
\rho_f=(A_i\otimes B_j)\,
\rho_r
\,(A_i^\dagger\otimes B_j^\dagger).
\end{equation}
The mean success probability can be calculated in accordance with Eq.~\eqref{eq:prob}.
\begin{equation}
p(r)=\frac{1}{9}
\sum\limits_{i,j=1}^{3}
\sum_{\xi\in\mathcal{S}_{ij}}
\Tr{\rho_f}
{\ketbra{\xi}{\xi}}
\label{eq:prob_rel}
\end{equation}
Since $\rho_f$ depends on $r$, mean success probability also depends on $r$.

%%%%%%%%%%%%%%%%%%%%%%%%%%%%%%%%%%%%%%%%%%%%%%%%%%%%%%%%%%%%%%%%%%%%%%%%%%%%%%%%
\section{Results and discussion}\label{sec:examples}
%%%%%%%%%%%%%%%%%%%%%%%%%%%%%%%%%%%%%%%%%%%%%%%%%%%%%%%%%%%%%%%%%%%%%%%%%%%%%%%%
Using Equations~\eqref{eq:rho-r}, \eqref{eq:game-final-noisy} and
\eqref{eq:prob_rel} we obtain the following formula for the probability of
winning the game as a function of the dimensionless acceleration:
\begin{equation}
p(r) = \frac19 \left( 6 + 2\cos 2r + \cos 4r \right).
\end{equation}

\begin{figure}[!h]
\centering\includegraphics{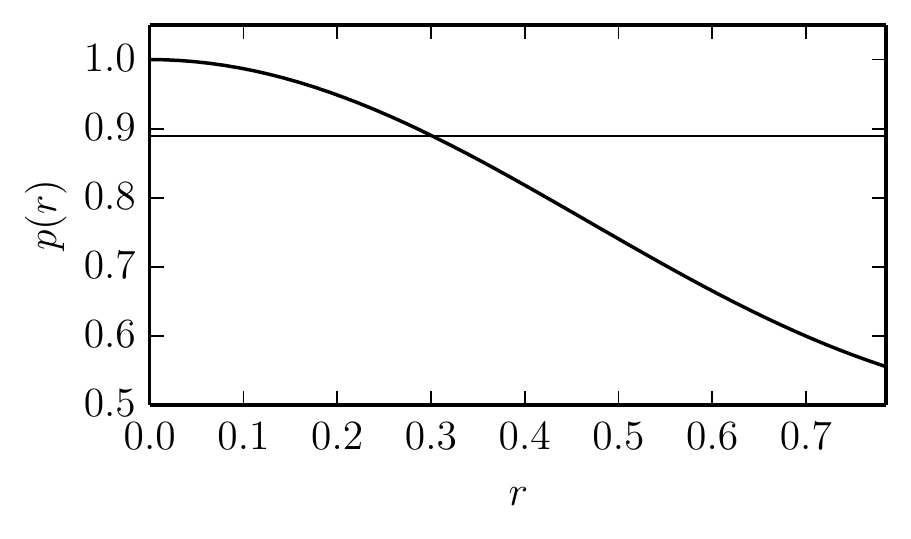}
\caption{Expected payoff of Alice and Bob as a function of Bob's acceleration. 
The thin horizontal line is the classical threshold. The lines intersect for $r 
= \tan ^{-1}\left(\sqrt{\frac{1}{3} \left(2 
\sqrt{7}-5\right)}\right)\approx
0.302171$.}\label{fig:unruh}
\end{figure}

The impact of constant acceleration on the probability of winning the game 
$p(r)$ is shown in Fig.~\ref{fig:unruh}. It shows that we can achieve 
probability of winning higher than the classical threshold for non-zero 
accelerations. For $r= \tan ^{-1}\left(\sqrt{\frac{1}{3} \left(2 
\sqrt{7}-5\right)}\right)\approx 0.302171$ players' probability of winning 
achieves the classical threshold of $8/9$, therefore the quantum strategy is 
indistinguishable from the classical one.
%%%%%%%%%%%%%%%%%%%%%%%%%%%%%%%%%%%%%%%%%%%%%%%%%%%%%%%%%%%%%%%%%%%%%%%%%%%%%%%%
\section{Summary}\label{sec:final}
%%%%%%%%%%%%%%%%%%%%%%%%%%%%%%%%%%%%%%%%%%%%%%%%%%%%%%%%%%%%%%%%%%%%%%%%%%%%%%%%
We studied the quantum Magic Square game in a relativistic setup. We introduced 
formalism which takes into account the influence of the Unruh effect on the 
outcome of the game. Obtained results show that it is possible to achieve 
pseudo-telepathy for non-zero accelerations.
%%%%%%%%%%%%%%%%%%%%%%%%%%%%%%%%%%%%%%%%%%%%%%%%%%%%%%%%%%%%%%%%%%%%%%%%%%%%%%%%
\section*{Acknowledgements}
%%%%%%%%%%%%%%%%%%%%%%%%%%%%%%%%%%%%%%%%%%%%%%%%%%%%%%%%%%%%%%%%%%%%%%%%%%%%%%%%
Work by {\L}P was supported by the Polish Ministry of Science and Higher
Education under the project number IP2012 051272.
Work by PG was supported by the Polish National Science Centre (NCN)
under the grant number N N516 481840.

\bibliography{relativistic_telepathy}
\bibliographystyle{unsrt}
\end{document}